\documentstyle[psfig]{caosp}
\begin{document}
\htitle{The magnetic variability of the star HD14437}
\hauthor{I.~I.~Romanyuk, D.~O.~Kudryavtsev}
\title{The magnetic variability of the star HD~14437}
\author{I.~I.~Romanyuk \and D.~O.~Kudryavtsev}
\institute{Special Astrophysical Observatory, Nizhnij Arkhyz,
Karachai--Circassian Republic, Russia, 357147}
\date{\today}
\maketitle
\begin{abstract}
The magnetic field of the CP star HD~14437
was discovered by Glagolevskij et al. (1985)
using the 6-m telescope of the Special Astrophysical Observatory.
No polarity changes have been found during 2 years of observations.
A very long ($>$~3--4 years) period of rotation was proposed to
explain the measurements. To check this hypothesis, we made
a new series of magnetic field observations for this star 10 years later with 
the 6-m telescope. The polarity of the longitudinal magnetic field is still
negative and has not shown any change during more than 10 years of
observations. We found the most probable periods to be several days.
It means that we have observed a star where the magnetic and rotation axes
are inclined at a small angle, and the negative pole is not far from the line
of sight.
\end{abstract}

\section{Observations and data reduction}
HD~14437 is a poorly studied peculiar A--star with a magnitude of $7^m.4$,
effective temperature $T_{\rm eff}=10800$ (Glagolevskij, 1995).
Observations of the 1980s with the 6-m telescope showed
a variability of the longitudinal magnetic field
with an amplitude of about 2 kG and a constant negative sign.
The description of the observations and data reductions
are presented by Glagolevskij et al. (1985) in more details.

In this paper we present new observations of HD~14437 which
were carried out in 1996--97 using the CCD detector
(Chuntonov \& Glagolevskij, 1997).
We have observed the spectra of the star on the 6-m
telescope with a spectral resolution $R=30000$. The context NICE
(Knyazev \& Shergin, 1995) in the MIDAS
system was used. The new and old observational data are presented in
Table~\ref{observations} (JD~2444655--2445900: with
photographic plates, JD~2449555 and JD~2449556: with the hydrogen lines
magnetometer, from JD~2450000 on: with the CCD detector).

We have carried out the line identification using
the Vienna Astrophysical Line Data--Base (VALD) (Piskunov et al., 1995)
and Moore's tables (Moore, 1945). In view of determining the individual
Land\'e factor of each spectral line used in our measurements,
we have identified the lines in the following spectral ranges:
$\lambda$5955--6385 and $\lambda$4460--4680.
Si and Cr are observed to be overabundant. We have found
an underabundance of O by the lack of the
lines OI~$\lambda$6155.97, 6556.78 and 6158.18 which are usually strong
at spectral class A2.

We have determined the effective magnetic field using Babcock's standard
formula with the individual Land\'e factor~$z$ for each line.
The measurements are given in Table~\ref{observations}.

\begin{table}[t]
\small
\begin{center}
\caption{The observations of the effective magnetic field.}
\label{observations}
\begin{tabular}{lr|lr}
\hline\hline
JD 2400000&$B_{\rm e}\pm\sigma$,G&JD 2400000&$B_{\rm e}\pm\sigma$,G\\
\hline
44655.208&$-1620\pm170$& 50415.240&$-1940\pm180$\\
44656.308&$-1030\pm170$& 50415.269&$-2570\pm180$\\
44659.188&$-440\pm170$&  50499.196&$-2650\pm260$\\
44660.196&$-800\pm140$&  50499.221&$-2600\pm160$\\
44860.541&$-2280\pm110$& 50500.161&$-2460\pm220$\\
45303.354&$-1230\pm120$& 50500.188&$-2050\pm250$\\
45303.362&$-1020\pm110$& 50617.507&$-1820\pm260$\\
45476.541&$-2080\pm280$& 50643.521&$-1360\pm110$\\
45900.493&$-2040\pm130$& 50705.451&$-1330\pm210$\\
49555.472&$-1340\pm380$*&50706.410&$-960\pm110$\\
49556.420&$-1710\pm460$*&50707.412&$-830\pm100$\\
50413.523&$-2040\pm300$& 50709.602&$-1370\pm110$\\
50414.125&$-2310\pm280$& 50710.432&$-2110\pm120$\\
50415.217&$-2460\pm270$&&\\
\hline\hline
\multicolumn{4}{l}{* --- observations with the magnetometer}
\end{tabular}
\end{center}
\end{table}

The effective magnetic field has preserved its sign for about ten years.
The suggestion about the long period of the star contradicts our data,
because we have observed the longitudinal component of the magnetic field
to undergo significant changes within a few days only. So, we can draw the
conclusion that the axis of rotation, the magnetic axis and the line of
sight are inclined at small angles.

\section{The period determination}

Having the observational data of the effective magnetic field $B_{\rm e}$,
we tried to find the rotational period of HD 14437.
It has been determined by Yurkevich's method in the interval of 0--100 days.
We have found three groups of probable periods: 13, 16 and 30 days.
The best fitted variability curve of $B_{\rm e}$ was derived for the period
of $16^{\rm d}.473$.

\begin{figure}[hbt]
\centerline{
\psfig{figure=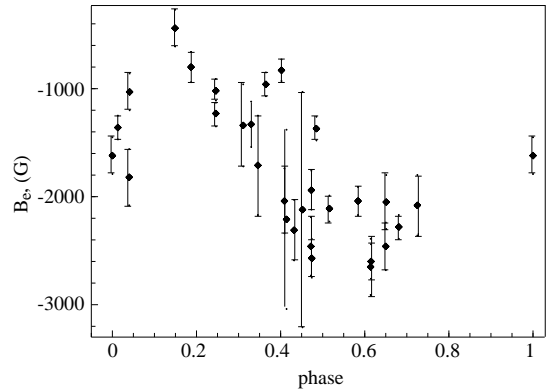,height=5.5cm}}
\caption{Variation of the effective magnetic field $B_{\rm e}$ for
$P=26^{\rm d}.732$.}
\label{curv26}
\end{figure}

Independently, a search for the period has been performed by Mathys et al. 
(1997) on the basis of the surface field $B_{\rm s}$.
They have found three groups of probable periods: 15, 30 and 350 days.
Later, Wade (1997) proposed another period,
close to $26^{\rm d}.732$, which he has calculated using 100 points of
Hipparcos photometry. The variability curve of the effective magnetic field
for $26^{\rm d}.732$ looks somewhat worse than for $16^{\rm d}.473$ because of
greater dispersion, nethertheless it remains quite acceptable. Besides,
determination of the period by photometry, with a much larger
number of homogeneous points, undoubtedly gives a more reliable result.
The $B_{\rm e}$ phase diagram for $P=26^{\rm d}.732$ is presented in 
Figure~\ref{curv26}.

\begin{acknowledgements}
We would like to thank V.~G.~Elkin for his help in observations
and discussions of the results, and G.~Wade for the information about
the 26 days period of Hipparcos photometry. This work was partly
supported by a RFBR grant 97--02--16247.
\end{acknowledgements}


\begin{thebibliography}{}
\inproceedings{Chuntonov, G.A., Glagolevskij, Yu.V.}{1997}
{Stellar magnetic fields}{Yu.V.~Glagolevskij \and I.I.~Romanyuk}{}{Moscow}{225}
\article{Glagolevskij, Yu.V., Bychkov, V.D., Romanyuk, I.I.,
 Chunakova, N.M.}{1985}{Astrofiz.~Issled.~(Izv.~SAO)}{19}{28}
\article{Glagolevskij, Yu.V.}{1995}{Bull.~of Spec.~Astrophys.~Obs.}{38}{}
\article{Knyazev, A.Yu., Shergin, V.S.}{1995}{SAO Technical report}{239}{}
\article{Mathys, G., Hubrig, S., Landstreet, J.D., Lanz, T., Manfroid, J.}
{1997}{\aaa}{123}{353}
\book{Moore, Ch.E.}{1945}{A multiplet table of astrophysical
 interest}{}{Princeton, New~Jersey}
\article{Piskunov, N.E., Kupka, F., Ryabchikova, T.A., Weiss, W.W.,
Jeffery, C.S. }{1995}{\aaas}{112}{525}
\bibitem{} Wade, G.: 1997, private comm.
\end{thebibliography}
\end{document}